\documentclass [twocolumn,showpacs,amsmath,amssymb,prl,supercriptaddress,floatfix]{revtex4}
\usepackage{graphicx}
\usepackage{dcolumn}
\usepackage{bm}
\begin{document}

\title{Negative Energy Resonances of Bosons in a Magnetic Quadrupole Trap}
\date{\today}
\author{Shahpoor Saeidian}
\email[]{s_saeid@physi.uni-heidelberg.de}
\affiliation{Physikalisches Institut, Universit\"at Heidelberg, Philosophenweg 12,
69120 Heidelberg, Germany}%
\author{Igor Lesanovsky}
\email[]{igor@iesl.forth.gr}
\affiliation{IESL - Institute of Electronic Structure and Laser,
FORTH - Foundation for Research and Technology,
P.O. Box 1527,GR-71110 Heraklion, Crete, Greece}
\author{Peter Schmelcher}
\email[]{Peter.Schmelcher@pci.uni-heidelberg.de}
\affiliation{Physikalisches Institut, Universit\"at Heidelberg,
Philosophenweg 12, 69120 Heidelberg, Germany}%
\affiliation{Theoretische Chemie, Institut f\"ur Physikalische Chemie,
Universit\"at Heidelberg, INF 229, 69120 Heidelberg, Germany}%
\date{\today}
\begin{abstract}\label{txt:abstract}
We investigate resonances of spin 1 bosons in a three-dimensional
magnetic quadrupole field. Complementary to the well-known positive energy
resonances it is shown that there exist short-lived, i.e.  broad, negative energy resonances.
The latter are characterized by an atomic spin that is aligned antiparallel to the
local magnetic field direction.
In contrast to the positive energy resonances the lifetimes of the negative energy
resonances decreases with increasing total magnetic quantum number.
We derive a mapping of the two branches of the spectrum.
\end{abstract}

\maketitle 

Magnetic traps represent a very powerful and versatile tool to study ultracold atomic
(or molecular) matter. In particular they are very well-suited for miniaturization via
e.g. so-called atomchips \cite{Folman2002,Reichel2002,Fortagh2005}
and consequently allow a processing of matter 
waves at the micro or even nanoscale. The basic idea behind the control of ultracold atoms
via inhomogeneous magnetic fields is very simple: The neutral atoms couple to the magnetic
field via their magnetic moment. Assuming strong hyperfine interactions (compared to the
magnetic interactions) the latter is due to the total angular momentum being composed of 
the electronic and nuclear angular momentum. Traditionally an adiabatic approximation,
which reduces the vector coupling of the magnetic moment to the field to a potential
term proportional to the magnitude of the field, is then employed. However, in order to
understand the basic physics of (individual) atoms in magnetic traps it is likewise necessary
to study the quantum dynamics of the Hamiltonian for the case of the vector coupling.
A number of investigations in this direction have been performed in the past.

Firstly fundamental configurations of static fields for the trapping of ground state
atoms have been explored in the late eighties and early nineties
\cite{Ketterle92,Bergeman87}. Planar current geometries for microscopic magnetic traps
were investigated in ref.\cite{Weinstein95}. Specifically the magnetic quadrupole field
\cite{Bergeman}, the wire trap \cite{Berg-Sorensen} and the magnetic guide as well as
the Ioffe trap \cite{Hinds1,Hinds2,Potvliege} have been addressed. Except for the
wire trap none of these configurations allow for bound states and we observe exclusively
a spectrum of resonance states. Hinds and Eberlein \cite{Hinds1,Hinds2} analyze the spectrum
and decay rates of resonances of particles with spin $\frac{1}{2}$ and $1$ 
in a magnetic guide by determining the phase shift of scattered waves.
Potvlieg\'{e} and Zehnle \cite{Potvliege} employ the complex scaling method
to obtain the widths and positions of the resonance states.
More recently Lesanovsky and Schmelcher \cite{Lesanovsky1} studied the dynamics of neutral
spin-$\frac{1}{2}$- and spin-$1$-particles in a 3D magnetic quadrupole field, equally
to ref.\cite{Bergeman}, but perform a much more extensive compilation of the properties
of the resonances including the derivation of an effective scalar Schr\"odinger equation
to describe and understand long-lived states possessing large angular momenta.

In the present work we investigate a new class of short-lived resonances 
possessing negative energies for the case of the 3D magnetic quadrupole field.
These resonances have obviously been overlooked in previous
investigations and originate from a fundamental symmetry of the underlying Hamiltonian.
Their interpretation in terms of spin anti-aligned transient states in 
the trap is outlined.

The Hamiltonian describing the motion of a point-like particle of
mass M and magnetic moment $\boldsymbol{\mu}$ in a 3D magnetic quadrupole field
${\bf{B}}= b(x,y,-2z)$ reads $H = \frac{\bf{p}^{2}}{2M} -
\boldsymbol{\mu}\cdot\bf{B}$. For a spin-S-particle with magnetic moment
$\boldsymbol{\mu}=-\frac{g}{2}\bf{S}$, one obtains in atomic units
\begin{equation}
H = \frac{1}{2M}[{\bf{p}}^{2} + bgM(xS_{x} + yS_{y} -2zS_{z})]
\end{equation}
Here $g$ is the $g$-factor of the particle.  Performing the scale
transformation $\bar{x}_{i}=(bgM)^{1/3}x_{i}$ and
$\bar{p}_{i}=(bgM)^{-1/3}p_{i}$ and omitting the bars one obtains
\begin{equation}\label{transHamil}
M(bgM)^{-2/3}H \rightarrow H = \frac{1}{2}({\bf{p}}^{2} + xS_{x} +
yS_{y} -2zS_{z})
\end{equation}
Therefore the energy level spacing scales according to
$\frac{1}{M}(bgM)^{2/3}$.  The scaled Hamiltonian ($\hbar = 1$) in spherical 
coordinates reads
\begin{equation}\label{spher.hamil.}
H = \frac{1}{2}(-\frac{\partial^{2}}{\partial r^{2}}
-\frac{2}{r}\frac{\partial}{\partial r} +\frac{L^{2}}{r^{2}}
+r\sin \theta K -2 r\cos \theta S_{z})
\end{equation}
Here $K = \cos\varphi S_{x} + \sin\varphi S_{y}$. For a spin-1-particle,
which is the case we focus on in the following except explicitly stated otherwise,
we have
\begin{displaymath}
K = \frac{1}{\sqrt{2}}\left(%
\begin{array}{ccc}
  0 & e^{-i\varphi} & 0 \\
  e^{i\varphi} & 0 & e^{-i\varphi} \\
  0 & e^{i\varphi} & 0 \\
\end{array}%
\right)\quad\mathrm{and}\quad S_{z} = \left(%
\begin{array}{ccc}
  1 & 0 & 0 \\
  0 & 0 & 0 \\
  0 & 0 & -1 \\
\end{array}%
\right)
\end{displaymath}

\begin{figure}[h]
$\begin{array}{c}
{\includegraphics[height=5cm,width=7.5cm]{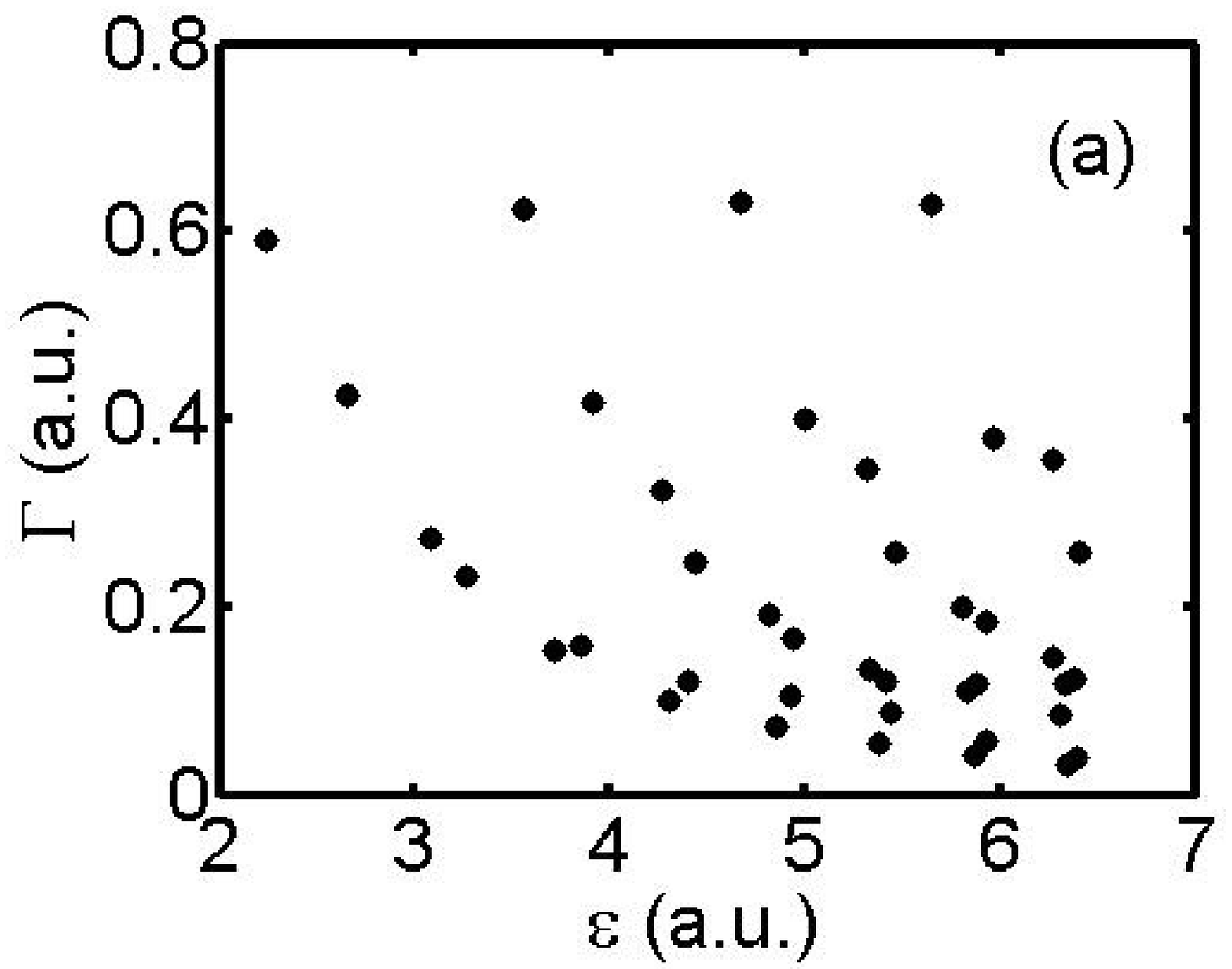}}\\
{  }\\
{\includegraphics[height=5cm,width=7.5cm]{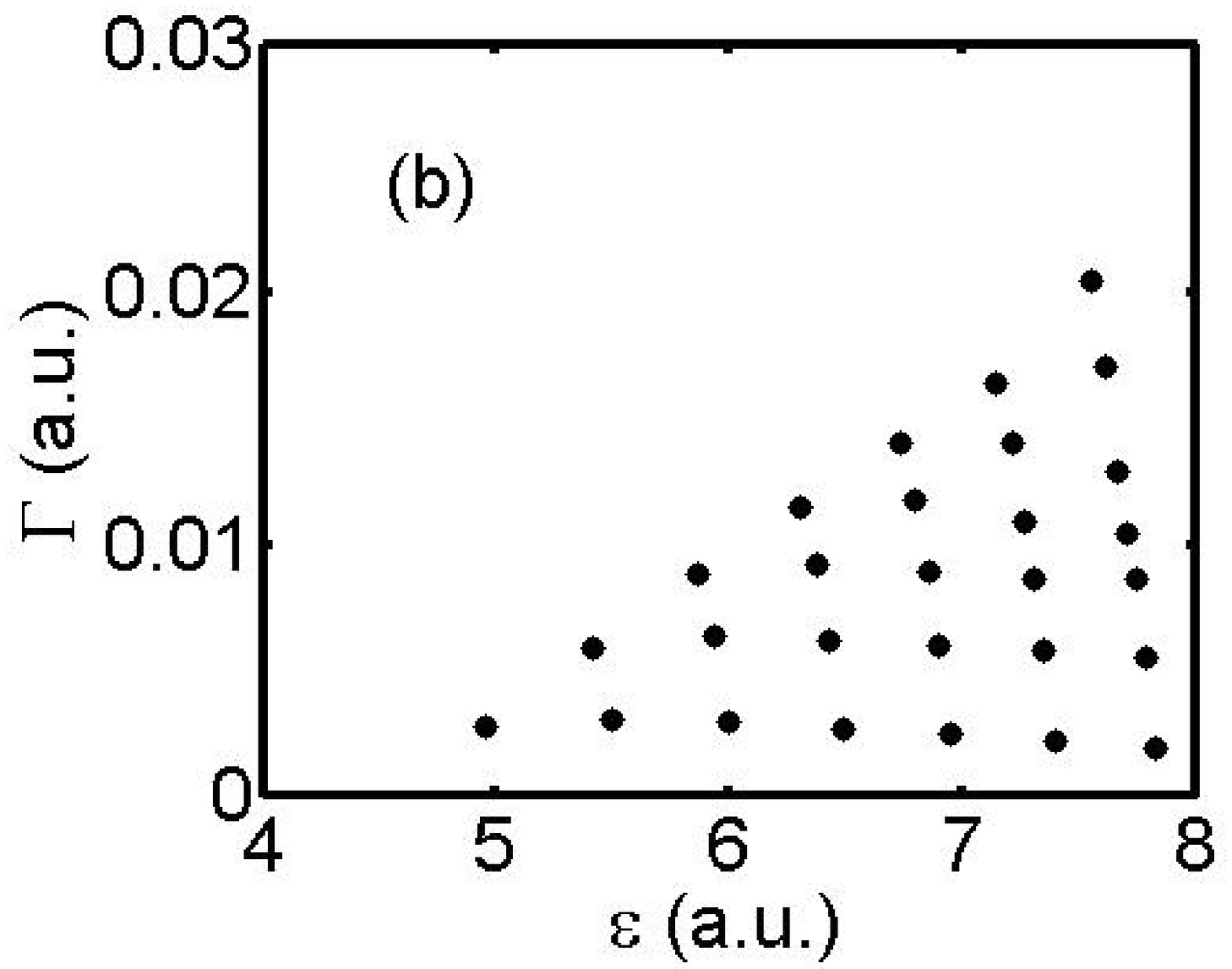}}\\
\end{array}$
\caption{Decay width and energies of the resonances possessing positive energies
for (a) $m=1$ and (b) $m=10$}
\label{positive energy resonances}
\end{figure}
By analyzing the symmetry properties of the
resulting Hamiltonian one obtains 15 discrete symmetries
(see ref.\cite{Lesanovsky1}).  Apart from the discrete symmetries
there is a continuous symmetry generated by $J_{z} = L_{z} +
S_{z}$ which is the $z$-component of the total angular momentum.  This is
a consequence of the rotational invariance of the system around
the $z$-axis of the coordinate system.  Due to $[J_{z}, H] = 0$
one can find energy eigenfunctions which are simultaneously
eigenfunctions of $J_{z}$.  For a spin-S-particle they read in the
spatial representation
\begin{equation}
| m \rangle^{(s)} = \sum_{m_{s} = -s}^{s}C_{m_{s}}e^{i(m -
m_{s})\varphi} | m_{s}\rangle
\end{equation}
with $\sum_{m_{s}} |C_{m_{s}}|^{2} = 1$ where $|m_s \rangle$ are the
spin eigenfunctions with respect to $S_z$.

Exploiting the discrete symmetries of the system, we find a
twofold degeneracy in the energy spectrum of the
Hamiltonian (\ref{transHamil}) for $m \neq 0$ where $m$ is the
quantum number of $J_{z}$.  The corresponding eigenstates of the
degenerate pair lie in $m$- and ($-m$)-subspaces and can
be identified with $|E, m\rangle$ and $|E, -m\rangle$.  Of course for $m =
0$ there is no degeneracy.

We are interested in the resonances of the
Hamiltonian (\ref{spher.hamil.}) which can be described by a
corresponding wave function localized in space (at $t = 0$). The
time evolution is given by
\begin{equation}
\psi _{R}(t) =  e^{-i\frac{E}{\hbar}t}\psi _{R}(0)
\end{equation}
where $E$ is complex
\begin{equation}\label{complexE}
E = \epsilon - i\frac{\Gamma}{2}
\end{equation}
$\epsilon $ and $\Gamma $ are the energy and decay width, respectively.
The imaginary part $-i\frac{\Gamma}{2}$, causes the resonances to decay with
a lifetime $\tau = \Gamma^{-1}$. 

In order to investigate the resonances we
employ the method of complex scaling (see ref.\cite{Moiseyev} and references therein).  For the computational details we refer the reader to ref.\cite{Lesanovsky1}.

We shall see below that the spectrum consists
of two well-separated parts: One set of resonances localized in the
negative energy region with short lifetimes and a second set
localized in the positive energy domain where the lifetimes are much
larger and can, at least in principle, become arbitrarily long. 
The positive energy resonances have already been investigated in several works
(see refs.\cite{Bergeman,Lesanovsky1} and references therein) whereas,
to our knowledge, no negative energy resonances have been reported up to date.

\begin{figure}[h]
$\begin{array}{c}
{\includegraphics[height=5cm,width=7.5cm]{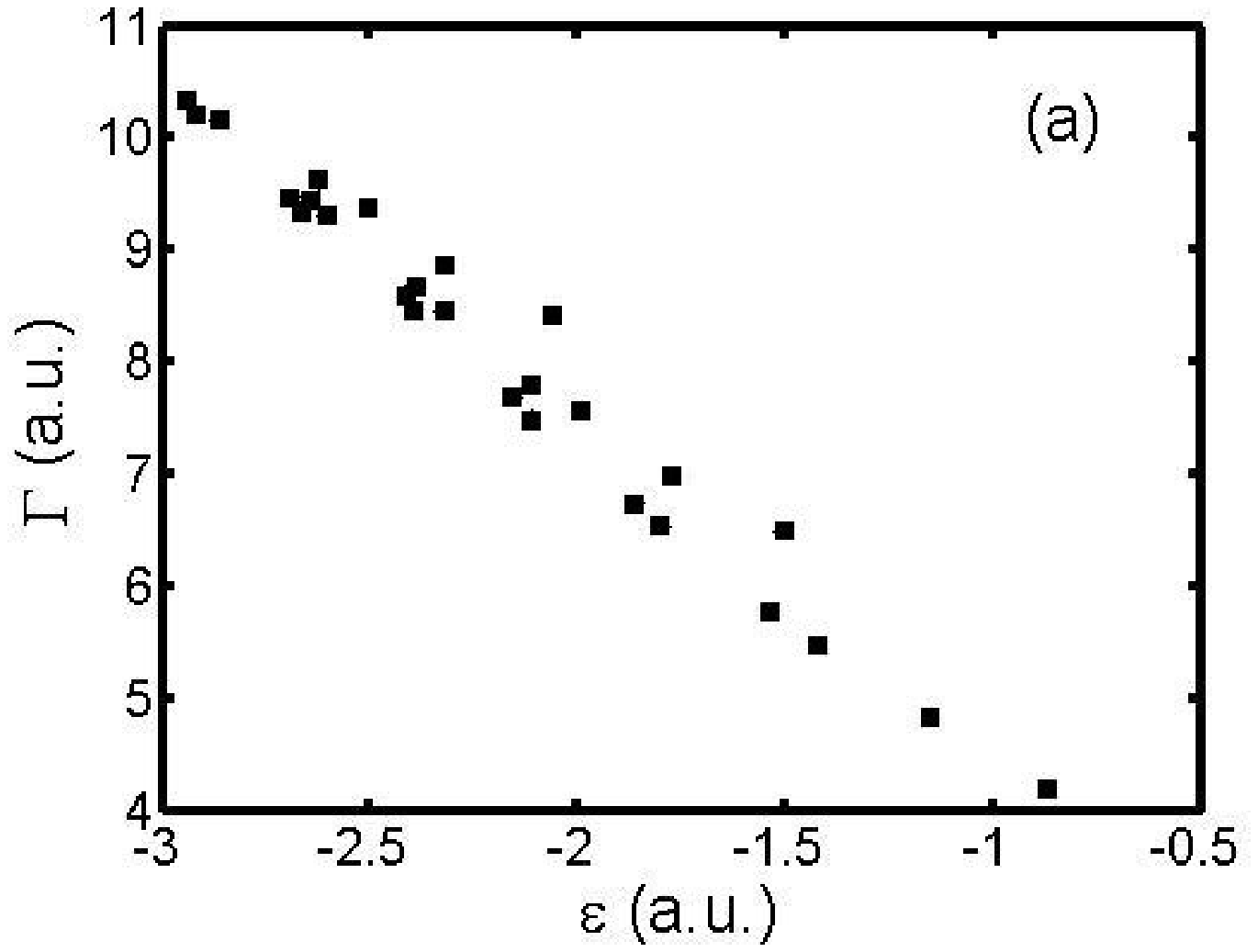}}\\
{  }\\
{\includegraphics[height=5cm,width=7.5cm]{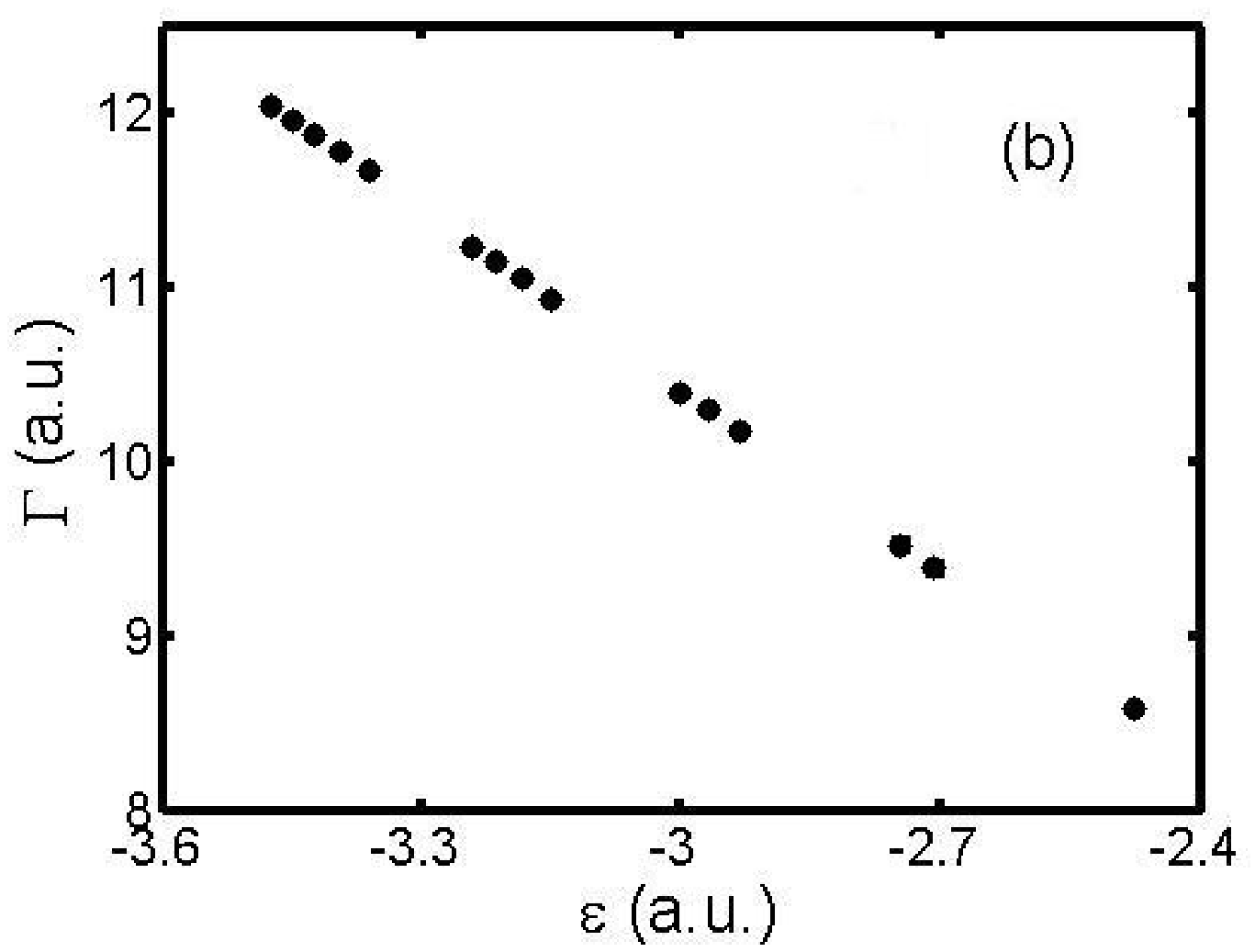}}\\
\end{array}$
\caption{Decay width and energies of the resonances possessing negative energies
for (a) $m=1$ and (b) $m=10$}
\label{negative energy resonances}
\end{figure}

In Fig.\ref{positive energy resonances} we present the energies and 
decay widths for positive energy resonances for a spin-1-particle.  For the
total angular momentum $m=1$ one observes the resonances to cover
the area of a right triangle in the $\epsilon-\Gamma$
plane. They are located on lines with similar negative slopes.
This pattern becomes increasingly disturbed
when considering resonances with higher energies and small decay widths.
For higher $m$ values (see figure \ref{positive energy resonances}(b) for
$m=10$) we observe a very regular pattern for their distribution 
in the $\epsilon-\Gamma$ plane, still with a triangular boundary, 
but now the resonances are located on lines possessing a positive slope 
(for a discussion of the origin of this regular pattern see
ref.\cite{Lesanovsky1}). The decay width decreases exponentially
with increasing value $m$ for the angular momentum.
The larger the angular momentum of a state the farther away it is located
from the center of the trap which is the zero of the magnetic field.
In the vicinity of the center spin-flip transition from e.g. bound to
unbound states take place. With increasing angular momentum the resonances
probe less and less of this central region and become therefore increasingly 
stable.

Fig.\ref{negative energy resonances} shows the energies and decay widths for 
the negative energy resonances.  One immediately notices that they
are arranged somewhat similarly to the ones with positive energy
but on lines with a {\it{negative slope}}.
The pattern of their distribution becomes more regular for higher $m$ values.
Unlike the positive energy resonances their decay width 
increases with increasing angular momentum.  States with a larger angular 
momentum experience a stronger magnetic field and, as we shall show below,
the magnetic moment of the atom is parallel to the local direction 
of the field in case of a negative energy resonance.
Consequently the atom feels an increasingly repelling force with respect to the
trap center if the total angular momentum increases. The latter leads 
to correspondingly broader resonances.

The knowledge of the resonance eigenfunctions of the complex-scaled
Hamiltonian enables us to calculate corresponding expectation values of observables
within the generalized inner $c-$ product being complex valued.
The real part represents the average value, whereas the imaginary part
can be interpreted as the uncertainty of our observable in a measurement
when the system is prepared in the corresponding resonance state (for more details see \cite{Moiseyev}).

\begin{figure}[h]
$\begin{array}{c}
{\includegraphics[height=5cm,width=7.5cm]{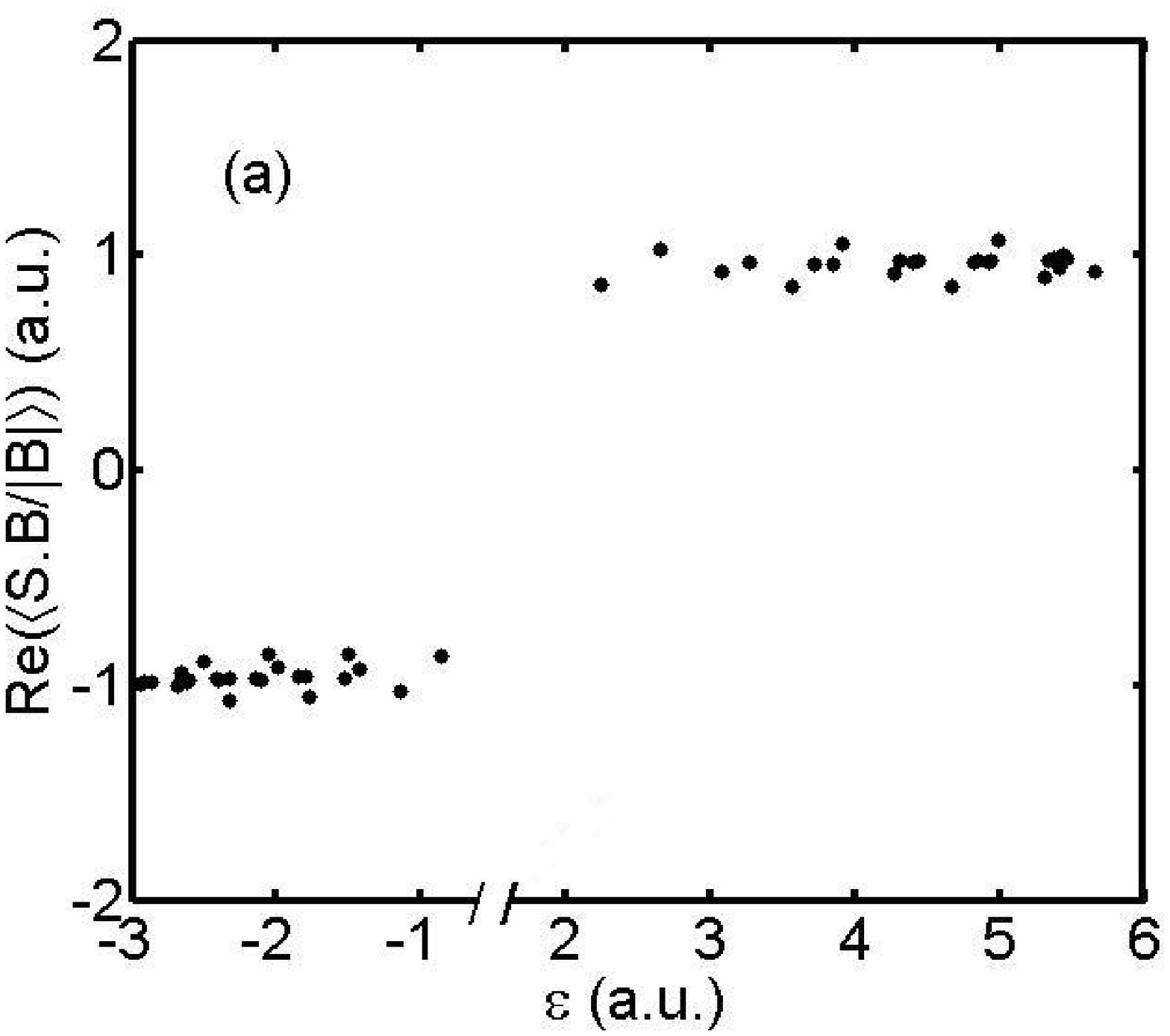}}\\
{  }\\
{\includegraphics[height=5cm,width=7.5cm]{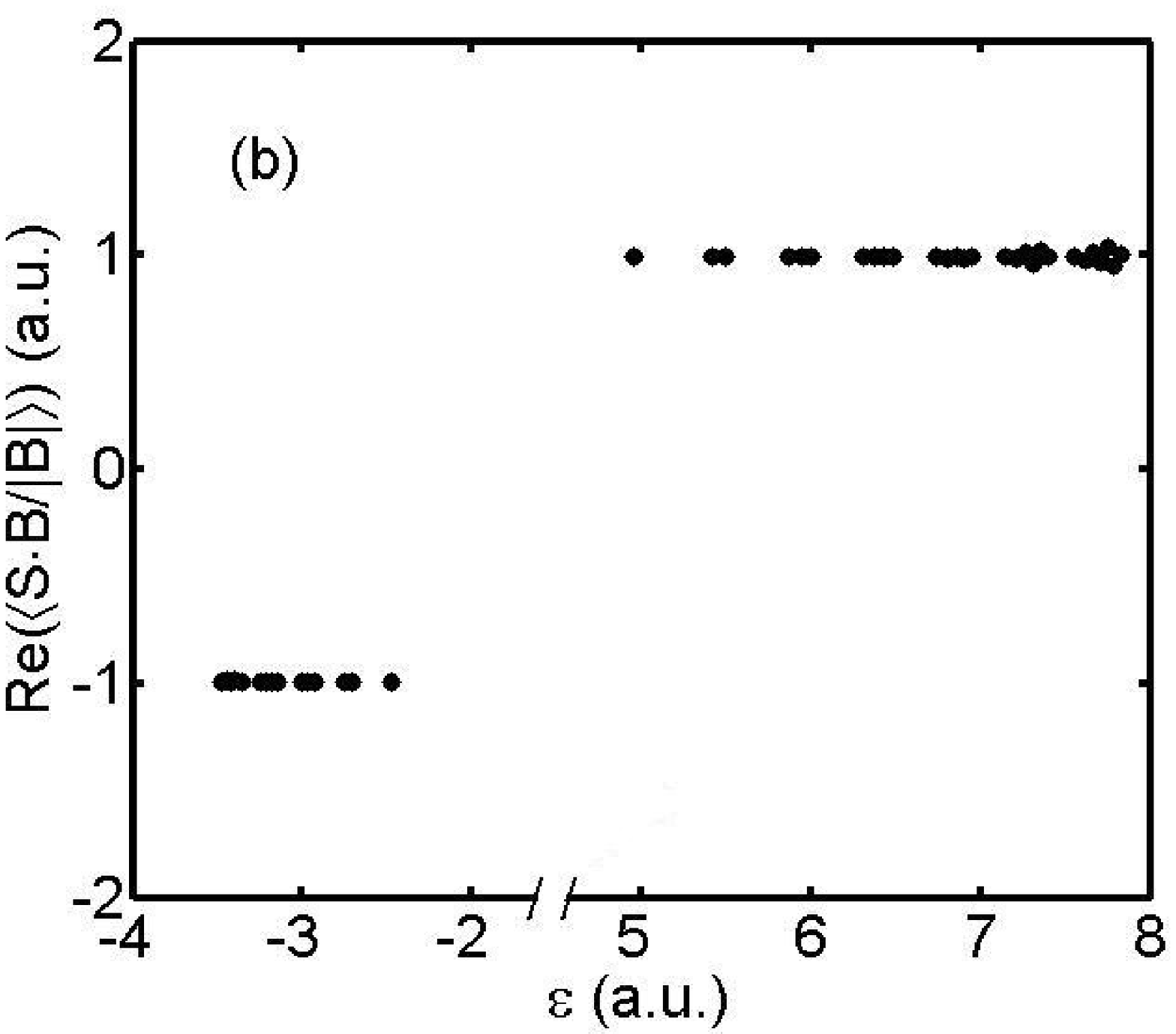}}\\
\end{array}$
\caption{Expectation values of the spin component along the local direction of the
magnetic field for the positive and negative energy resonances for (a) $m=1$ and (b) $m=10$.}
\label{spin component}
\end{figure}

In Fig.\ref{spin component} we present the expectation value 
$Re(\langle {\bf{S}} \cdot {\bf{B}} / |B| \rangle)$
of the spin component which points along the
local direction of the field as a function of the energy for the two cases $m=1$ and 
$m=10$.  For the negative energy resonances this value is approximately $-1$ 
indicating that the spin is aligned opposite to the local direction of the magnetic field and
the magnetic moment is in the same direction as the field (we have assumed
$g>0$).  For positive energy resonances the spin is parallel to the
local field and the magnetic moment is antiparallel. 

\begin{figure}[h]
$\begin{array}{c}
{\includegraphics[height=5cm,width=7.5cm]{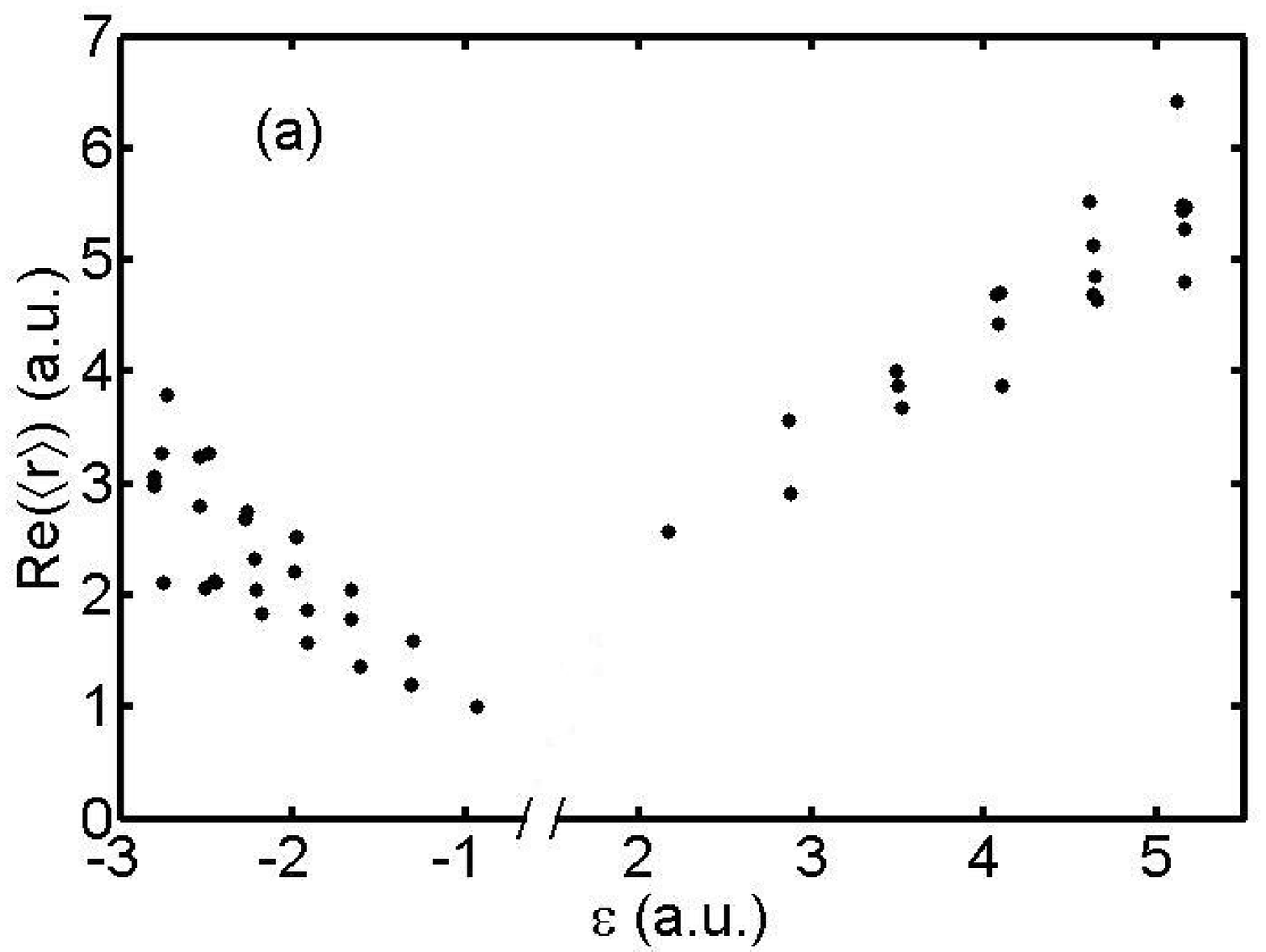}}\\
{  }\\
{\includegraphics[height=5cm,width=7.5cm]{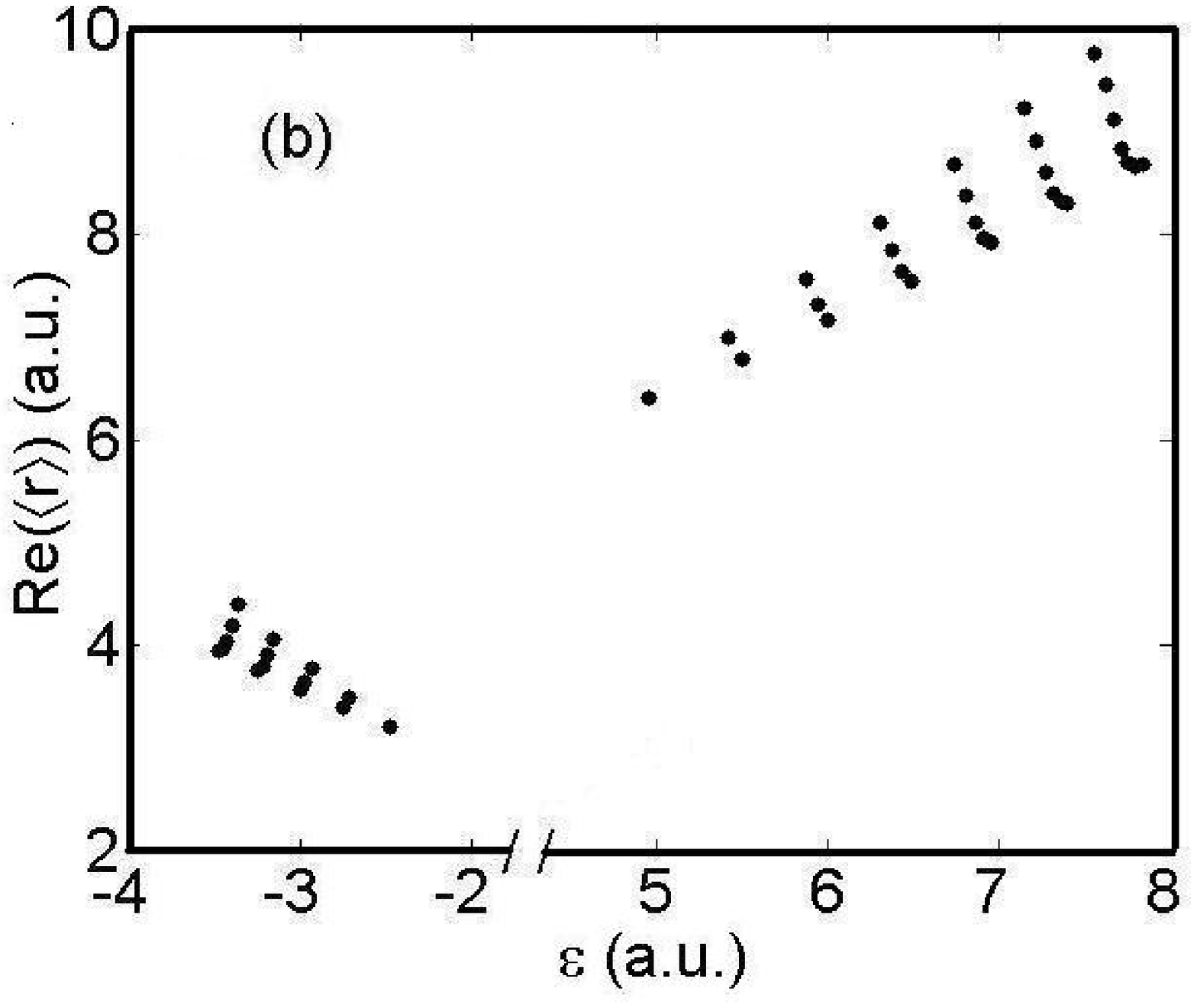}}\\
\end{array}$
\caption{Average of the radial coordinate $r$ as a function of the energy for both positive
and negative energy resonances for (a) $m=0$ (b) $m=10$.} 
\label{r average}
\end{figure}

It is instructive to consider the average value of the radial coordinate $r$
as a function of the energy (see Fig.\ref{r average}) . The resonances
are distributed in an area possessing the shape of a triangle. Again this pattern becomes increasingly
regular, when considering higher angular momenta (see fig.\ref{r average}(b)). 
$Re(\langle r\rangle)$ increases with increasing absolute value of the energy
which is in agreement with our above argumentation with respect to the localization
of the resonances.

The results on the negative and positive energy resonances suggest an intimate
relationship among the two. Indeed due to a complex symmetry, which we shall derive
in the following, there exists a mapping of the two classes of resonances.
The Schr\"{o}dinger equation belonging to the Hamiltonian which results when applying
the complex scaling to the Hamiltonian (\ref{spher.hamil.}) reads
\begin{eqnarray}\label{Co.Sc.Sch.Eq}\nonumber
\frac{1}{2}(-e^{-i2\eta}\frac{\partial^{2}}{\partial
r^{2}}-e^{-i2\eta}\frac{2}{r}\frac{\partial}{\partial
r}+e^{-i2\eta}\frac{m^{2}}{r^{2}}+e^{i\eta}r\sin\theta
K
\end{eqnarray}
\begin{eqnarray}
-2e^{i\eta}r\cos\theta
S_{z} )\psi_{E,m}=E\psi_{E,m}
\end{eqnarray}
Taking the complex conjugate and multiplying by $e^{-i\frac{2\pi}{3}}$ one obtains
\begin{eqnarray}\nonumber
\frac{1}{2}(-e^{-i2(\pi /3-\eta)}\frac{\partial^{2}}{\partial
r^{2}}-e^{-i2(\pi /3-\eta)}\frac{2}{r}\frac{\partial}{\partial r}
\end{eqnarray}
\begin{eqnarray}\nonumber
+e^{-i2(\pi /3-\eta)}\frac{m^{2}}{r^{2}}
+e^{i(\pi/3-\eta)}e^{-i\pi}r\sin\theta K^{*}
\end{eqnarray}
\begin{eqnarray}
-2e^{i(\pi/3-\eta)}e^{-i\pi}r\cos\theta
S_{z}^{*})\psi_{E,m}^{*}=e^{-i\frac{2\pi}{3}}E^{*}\psi_{E,m}^{*}
\end{eqnarray}
By transforming
\begin{displaymath}
S_{i}\rightarrow \bar{S}_{i}=e^{-i\pi}S_{i}^{*}=-S_{i}^{*}
\end{displaymath}
\begin{equation}\label{transS}
\eta\rightarrow \bar{\eta}=\pi/3 - \eta
\end{equation}
one can show that
\begin{eqnarray}\nonumber
\frac{1}{2}(-e^{-i2\bar{\eta}}\frac{\partial^{2}}{\partial
r^{2}}-e^{-i2\bar{\eta}}\frac{2}{r}\frac{\partial}{\partial
r}+e^{-i2\bar{\eta}}\frac{m^{2}}{r^{2}}+e^{i\bar{\eta}}r\sin\theta \bar{K}
\end{eqnarray}
\begin{eqnarray}
-2e^{i\bar{\eta}}r\cos\theta \bar{S}_{z})\bar{\psi}_{\bar{E},m} = \bar{E} \bar{\psi}_{\bar{E},m}
\end{eqnarray}
which is unitarily equivalent to eq.(\ref{Co.Sc.Sch.Eq}) with 
the transformed energies and resonance wave functions
\begin{displaymath}
\nonumber
E \rightarrow \bar{E} = e^{-i\frac{2\pi}{3}}E^{*}\\ 
\end{displaymath}
\begin{displaymath}
\psi_{E,m}
\rightarrow\bar{\psi}_{\bar{E},m}(e^{i\bar{\eta}}r,
\theta ,\varphi ,m_{s})=\psi_{E,m}^{*}(e^{i\eta}r, \theta ,\varphi ,-m_{s})
\end{displaymath}

Therefore each resonance belonging to eq.(\ref{Co.Sc.Sch.Eq}) with eigenvalue $E$
possesses a counterpart i.e. a 'partner resonance' with the eigenvalue $\bar{E}= e^{-i2\pi / 3}E^{*} $.
We therefore have a mirror symmetry of the underlying Hamiltonian the corresponding mirror line being
placed at $-\pi / 3$. While the positive energy continuum rotates clockwise around its zero energy
threshold by $2\eta$ its image rotates anticlockwise by $2\eta$ starting from the line defined by
the polar angle $- \frac{2 \pi}{3}$. When a resonance is revealed at the angle $-2\eta_{0}$
in the positive energy domain, its image resonance for negative energies
is revealed at the polar angle position $-2\pi / 3 + 2\eta_{0}$.

Employing the transformation (\ref{transS}), the
expectation values of $r$ and of the spin-component along the local
direction of the magnetic field 
$\frac{\bf{S}\cdot\bf{B}}{|\bf{B}|}$ transforms as follows

\begin{displaymath}
\langle r\rangle \rightarrow e^{i\pi /3}\langle r\rangle^{*}
\end{displaymath}
\begin{displaymath}
\langle \frac{\bf{S}\cdot\bf{B}}{|\bf{B}|}\rangle \rightarrow
e^{i\pi }\langle
\frac{\bf{S}\cdot\bf{B}}{|\bf{B}|}\rangle^{*}=-\langle
\frac{\bf{S}\cdot\bf{B}}{|\bf{B}|}\rangle^{*}
\end{displaymath}
which are in good agreement with the results of our numerical calculations.
Finally we remark that although our study was performed for the specific case of a spin-1-particle
our conclusions hold for any boson with non-zero spin and also for fermions.

Let us conclude. We observe, to our knowledge, for the first time
negative energy resonances of spin-1-bosons in a 3D quadrupole magnetic trap.
The overall spectrum is arranged in two disconnected parts each of which contains exclusively resonances
with positive and negative energies. The latter are exclusively of short-lived character
and the spin is antiparallel to the local direction of the magnetic field while for the former
the spin is parallel to the field. As the total angular momentum of the boson
increases, the decay width of a negative energy
resonance state increases while for a positive energy resonance it decreases
(assuming a typical laboratory gradient field $5 T/m$ and the species $Rb-87$ 
the order of magnitude for the lifetimes of the negative energy resonances is mikroseconds).
A property of the complex scaled Hamiltonian has been established which allows to map
the two branches of the spectrum.

S.S. acknowledges financial support by the Ministry of Science, Research and Technology of Iran 
in the form of a scholarship.

\end{document}